\documentclass[epsf]{elsart}
\usepackage{graphicx}

\input epsf
\begin{document}
\begin{frontmatter}
\title{An effective Lagrangian description of
charged Higgs decays $H^+ \to W^+ \gamma, \, W^+Z$ and $ W^+h^0$}
\author[ifbuap]{J. L. D\'\i az--Cruz},
\author[ifbuap]{ J. Hern\' andez--S\' anchez},
\author[fcfmbuap]{J. J. Toscano}
\address[ifbuap]{Instituto de F\'{\i}sica, BUAP, Apartado Postal J-48,
72570 Puebla, Pue. M\'exico}
\address[fcfmbuap]{Facultad de Ciencias F\'\i sico Matem\'aticas, BUAP,
      Apartado Postal 1152, 72000 Puebla, Pue. M\'exico }

\begin{abstract}
Charged Higgs decays are discussed within an effective 
lagrangian extension of  the two-higgs doublet model, assuming new physics appearing in the Higgs sector of this model. Low-energy contraints  are used to imposse bounds on certain dimension--six operators that describe the modified charged Higgs interactions. These bounds are used then to study the decays $H^+ \to W^+\gamma, \, W^+Z$ and $ W^+h^0$, which can have branching ratios (BR) of order $10^{-5}$, $10^{-1}$ and $O(1)$,  respectively; these modes are thus  sensitive probes of the symmetries of the Higgs sector that could be tested at future colliders.
\end{abstract}
\end{frontmatter}

{\bf {1.- Introduction.}}
The scalar spectrum of many well motivated
extensions of the Standard Model (SM) include a charged Higgs state,
whose detection  at future colliders would constitute a clear evidence
of a Higgs sector beyond the minimal SM. In particular, the two-Higgs 
doublet model (THDM) has been extensively studied as a prototype of a
 non-minimal Higgs sector that includes a charged Higgs boson
 ($H^\pm$)~\cite{kanehunt}.
However, a definite test of the mechanism of electroweak symmetry breaking will require further studies of the complete Higgs spectrum. In particular, probing the properties of charged Higgs could help to find out whether it is indeed associated with
a weakly-interacting theory, as in the case of the popular 
minimal SUSY extension of the SM (MSSM) \cite{susyhix}, or to an strongly
interacting scenario, as in the technicolor models (or its
relatives top-color and top-condensate models) \cite{stronghix}.
Furthermore, these tests should also allow to probe the symmetries of the
Higgs potential, and to determine whether the charged Higgs belongs
to a weak-doublet or to some larger multiplet.

Decays of a charged Higgs boson have been studied
in the literature, including the radiative modes  into
$W^+\gamma, W^+Z$ \cite{hcdecay}, but mostly within the context
of the THDM,  or its MSSM incarnation. Charged Higgs production
at hadron colliders was studied long ago \cite{ldcysampay}, 
and recently more systematic calculations of production
process at  LHC have been presented \cite{newhcprod}.
Current bounds on charged Higgs mass can be obtained at
Tevatron, by studying the top decay $t\to bH^+$, which
already eliminates some region of parameter space \cite{LHCbounds}, 
whereas LEP bounds give 
approximately $m_{H^+} > 100$ GeV \cite{LEPbounds}.

The decays $H^+\to W^+\gamma$ and $H^+\to W^+Z$ may be quite sensitive to
new physics effects since they are loop-predicted within the THDM.
Indeed, the $W^+\gamma$ mode cannot be induced at tree level in any
renormalizable theory due to electromagnetic gauge invariance.
In turn, the absence of the $W^+Z$ mode at tree--level is a
feature of models that include only Higgs doublets; it can be induced
at this level only in models with Higgs triplets or higher representations,
though it is strongly constrained by the custodial symmetry $SU(2)_c$.
On the other hand, the decay $H^+\to W^+ h^0$ happens to be also
suppressed  for the THDM, when its parameters resemble 
the ones of the MSSM, specially for its decoupling limit,
 i.e. when $m_A>> m_Z$. Though suppressed, these decay modes deserve special attention because they can give valuable information about the underlying structure of the gauge and scalar sectors. Besides, these modes have a clear signature and could be detected at future Hadron colliders. The decay modes of a relatively light charged Higgs
boson, with mass of order of the Fermi scale, will depend on the
specific structure of a more fundamental theory that incorporates
new heavy fields. In this paper we will perform a general study
of these decays in a model--independent manner using the effective
Lagrangian technique, which is a well motivated scheme to parametrize the virtual effects of physics beyond a given theory, in our case the THDM.
We will assume that the spectrum of physical scalars predicted by the THDM
are relatively light ($m< O(1)$ TeV) and thus they can be specified within
a linear realization of the electroweak group.
This corresponds to the decoupling scenario, where the heavy fields
cannot affect dramatically the low energy process,
though they may have significant
contributions on those couplings that are absent or
highly  suppressed within the THDM.
The effective Lagrangian that we will use in this study
is a natural extension of the one given in \cite{BW} for the minimal SM,
which was extended in \cite{wudkamexico} to study rare decays of the
neutral CP--odd scalar predicted by the THDM.

The main goal of this paper is to study the decays of the charged
Higgs boson, both  within the THDM and its  effective Lagragian
extension, as possible probes of the Higgs sector.
Charged Higgs decays are first discussed within the THDM, which is extended by including higher-dimensional operators. Then, using low-energy data, e.g. the S,T,U parameters, we are able to impose bounds on certain dimension--six operators that also induce modifications to the charged Higgs interactions. We use these bounds to predict the branching ratio for the modes $H^+ \to W^+\gamma, W^+Z$ and $W^+h^0$, which can reach values that could be tested at future colliders.

{\bf{2.- Decays of the charged Higgs in the THDM.}}
One of the simplest models that predicts a charged Higgs is the THDM,
which includes two scalar doublets of equal hypercharge, namely
$\Phi_1=(\phi^+_1,\phi^0_1)$ and $\Phi_2=(\phi^+_2,\phi^0_2)$.
This is related  to the Higgs content used in the minimal SUSY
extension of the SM (MSSM). Besides the charged Higgs ($H^\pm$),
the spectrum that arises in the THDM includes two neutral CP-even
states ($h^0,H^0$, with $m_h<m_H$), as well as a neutral CP-odd
state ($A^0$). The most general Higgs potential that includes a
softly-broken
discrete symmetry $\Phi_1 \to \Phi_1$ and $\Phi_2 \to -\Phi_2$,
is given by:
\begin{eqnarray}
V(\Phi_{1} , \Phi_{2}) & = & \mu_{1}^{2} \Phi_{1}^{\dag } \Phi_{1} +
\mu_{2}^{2} \Phi_{2}^{\dag } \Phi_{2} - (\mu_{12}^{2} \Phi_{1}^{\dag }
\Phi_{2}+ h.c. )
+ \lambda_{1} (\Phi_{1}^{\dag } \Phi_{1})^{2} \nonumber  \\
& + & \lambda_{2} ( \Phi_{2}^{\dag } \Phi_{2} )^{2} 
+  \lambda_{3} (\Phi_{1}^{\dag } \Phi_{1}) (\Phi_{2}^{\dag }\Phi_{2}) 
+ \lambda_{4} (\Phi_{1}^{\dag } \Phi_{2}) (\Phi_{2}^{\dag }\Phi_{1})
\nonumber \\
& + & \frac{1}{2 } [\lambda_{5} (\Phi_{1}^{\dag }\Phi_{2})^{2} +h.c]
\end{eqnarray}
Diagonalization of the resulting mass matrices gives the expression for
the charged Higgs mass-eigenstate:
$H^+=\cos\beta \phi^+_1 + \sin\beta \phi^+_2$,
where $\tan\beta(=v_2/v_1)$ denotes the ratio of v.e.v.'s
from each doublet. The charged Higgs mass is given by:
\begin{equation}
m^2_{H^{\pm}}= m^2_A+ \frac{2m^2_W}{g^2} (\lambda_5-\lambda_4)
\end{equation}
When $\lambda_5=\lambda_4$ we have $m_{H^+}=m_A$, which reflects the
underlying custodial symmetry of the Higgs potential.

The predictions for the charged Higgs decays that arise within
the THDM and beyond, can be interpreted as possible probes of 
the symmetries of the Higgs sector. For instance, if we
focus on the gauge interactions of the charged Higgs, then the
coupling $H^+W^-h^0$ is quite sensitive to the structure of the
covariant derivative, and could be one place where to look
for deviations from the minimal THDM (or SUSY) predictions.
This vertex could induce the decay $H^+ \to W^+h^0$, whenever it
is kinematically allowed; the corresponding decay width is given by:
\begin{eqnarray}
\Gamma(H^+\to W^+h^0)&=& 
\frac{g^{2}\lambda^{\frac{1}{2}} (m^2_{H^\pm},m^2_{w}, m^2_{h^o}) }
{64 \pi m^{3}_{H^{\pm}}} 
c^2_{\beta -\alpha} \nonumber \\
&\times& \left[ m_{w}^2- 2(m_{H^{\pm}}^2+ m^2_{h^o}) +
\frac{(m_{H^{\pm}}^2- m^2_{h^o})^2}{m^2_{w}}\right]
\end{eqnarray}
where $\lambda $ is the usual kinematic factor,
$ \lambda(a,b,c)= (a-b-c)^{2}-4bc$.
This decay  mode has been studied in the literature \cite{hcwhdetect},
where it is concluded that its detection at the coming large hadron
collider (LHC)
is feasible. Within the THDM  this decay is proportional to the factor
$c_{\beta -\alpha }^2$, which will determine the strength of
this decay. For instance,
within the MSSM, $c_{\beta -\alpha }^2\sim m_z^2/m_{A^o}^2 $,
which tends to be small for large values of $m_A$, except for
small regions of parameter space. Although the BR for 
this mode can be small in the THDM, new physics
could enhance it. 

Within the THDM, and other models which treat the charged Higgs as
an elementary field, the decay $H^+ \to W^+\gamma$ only arises at
the loop level, and tends to has a very small BR (typically smaller
than about $10^{-5}$), which could be considered as a generic feature
of an elementary Higgs. However, when $H^+ $ arises from a composite
model, the corresponding BR could be enhanced and reach detectable
levels \cite{composites}. Similarly, the decay $H^+ \to W^+Z$
arises at the loop-level in the THDM, but now for some regions of
parameters it could have a large BR, which is a remnant of non-decoupling effects present in the model. On the other hand, in
models with Higgs triplets, the decay $H^+ \to W^+Z$ can arise at
tree-level, as  a result of violations of the custodial symmetry,
which is related to the observed value $\rho \simeq 1 $. Thus
$H^+ \to W^+Z$ can also be used to study the symmetries of the
Higgs sector.

Other relevant decays of the charged Higgs boson are the modes into
fermion pairs, which include the decays $H^+ \to \tau\nu_\tau, c \bar{b}$,
and possibly into $t\bar{b}$. If the
charged Higgs is indeed associated with the Higgs mechanism, its
couplings to fermions should come from the Yukawa sector, and the
corresponding decays should have a larger BR for the modes involving the
heavier fermions. A very simple test of this could be
done through a comparison of the modes $H^+ \to \tau\nu_\tau$ and
$H^+ \to \mu\nu_\mu$, which should be quite different if the charged
Higgs comes as a remmant of the Higgs mechanism.

In order to discuss the new results of the following section,
and to present our notation and conventions, we found
convenient to discuss here the loop-decays of the charged Higgs in
the THDM. Although these decays have been partially discussed in
the literature, we have also opted to perform our own calculation
to verify previous results, for which we find complete agreement.
We have evaluated the corresponding amplitudes for both
$H^+ \to W^+\gamma$ and $H^+ \to W^+Z$, using dimensional
regularization, with the help of the programs Feyncalc~\cite{Feyncalc}
and the numerical package FF~ \cite{FFloop}. We shall not present
here all the detailed analytical expressions for the amplitudes,
which are obtained using the non-linear gauge, as described
previously in ref.\cite{ouregeh}; instead only their generic
form will be displayed. The complete list of diagrams encountered in the
calculation, and the full expressions for the results
will be presented elsewhere \cite{ournext} .

Using a non-linear gauge eliminates the
three-point vertices of the type $WVG$ (where $V$ represents
the neutral gauge bosons, and $G$ denotes the charged Goldstone boson),
which reduces considerably the number of diagrams, this helps to
simplify the calculation and
to verify the gauge invariance in the case of the $W\gamma$ mode.
Our result for the total amplitude of the decay
$H^+ \to W^+ V$ (with $V=\gamma,Z$) can be summarized as:
\begin{equation}
 \mathcal{M}_V = -\frac{ig  }{m_W} 
  \epsilon^{\mu} \epsilon^{\nu}
 \bigg [m^2_W F_V g_{\mu\nu} + G_V k_{1\mu}k_{2\nu} \nonumber \\
 +iH_V \epsilon_{\mu\nu\alpha\beta} k^{\alpha}_1k^{\beta}_2 \bigg ],
\end{equation}
where $F_V,G_V,H_V$ denote the contributions from
the loop graphs to the amplitudes, and
$\epsilon^{\mu}(k_1,\lambda_1)$ and $\epsilon^{\nu}(k_2,\lambda_2)$
represent the polarization vectors of the $W$ and $V$ gauge bosons,
respectively.

For the mode $H^+ \to W^+\gamma$ we have: $m^2_WF_V = -G_V k_1.k_2$,
and then the corresponding decay width is given by:
\begin{equation}
 \Gamma (H^+\to W^+\gamma) = \frac{g^2 m^3_{H^+}}{32\pi m^2_W}
       \lambda^{3/2}(1,w,0) [ |G_\gamma|^2 + |H_\gamma|^2]
\end{equation}
where $\lambda(x,w,z)= (x-w-z)^2-4wz$, and $w=m^2_W/m^2_{H^+}$, 
$z=0$.

On the other hand, the decay width for the mode $H^+ \to W^+Z$ 
is conveniently written as:
\begin{equation}
 \Gamma (H^+\to W^+Z) = \frac{ m_{H^+}}{16\pi}
       \lambda^{1/2}(1,w,z) [ |V_{TT}|^2 + |V_{LL}|^2]
\end{equation}
where   here 
$w=m^2_W/m^2_{H^+}$, $z=m^2_Z/m^2_{H^+}$ and the quantities
$|V_{TT}|^2$ and $|V_{LL}|^2$ are the separate contributions
of the longitudinaly and transversely polarized final W- and
Z-bosons. They are explicitly given by
\begin{eqnarray}
|V_{LL}|^{2} & = & \frac{g^{2}}{4z} \left|  (1-w-z)F_Z + 
                  \frac{\lambda (1,w,z) }{2w}  G_Z  \right| ^{2}  \\
|V_{TT}|^{2} & = & g^{2} \left[  2 w |F_Z|^{2}                   
+  \frac{ \lambda (1,w,z) }{2w} |H_Z|^{2} \right]  \nonumber 
\end{eqnarray} 

 To evaluate the branching
ratios we have used the expressions for the decay
widths of the tree-level modes, as appearing in ref. \cite{kanehunt}.
We have taken $m_t=175\;$ GeV, 
and the values for the electroweak parameters of the
table of particle properties \cite{Partdat}.
 We shall present the resulting BR for
the charged Higgs decays in Figs. 1-3 of the following
sections. For the moment we only mention the BR for these decays for the following three relevant scenarios, which assume $m_h=115$ GeV,

{\bf a) SUSY-like scenario.} Here we assume
an approximately degenerated spectrum  of heavy Higgs bosons, i.e.
$m_{H^+}\simeq m_H\simeq m_A$, and also $\alpha\simeq \beta-\pi/2$.
In this case the BR for the mode $W^+h^0$ is about
$2\times 10^{-2}$ ($ 7 \times 10^{-5}$) 
for $m_{H^+}=300$ GeV and $\tan\beta=7$ (30).
On the other hand, the BR for the mode $H^+ \to W^+Z$ is about
$10^{-3}$ ($3\times 10^{-5}$), whereas for $H^+ \to W^+\gamma$ is about
$2\times 10^{-6}$ ($ 10^{-7}$)   
for the same values of
parameters. In this scenario $H^+\to W^+h^0$ and $H^+\to W^+Z$ have similar BR's.

{\bf b) Non-decoupling scenario-A.} Now we take a large mass difference between $A^0$ and $H^+$, i.e. $m_{H^+}- m_A=300$ GeV, with $m_H \simeq m_{H^+}$, and also $\alpha\simeq \beta-\pi/2$. In this case we find that the mode $W^+h^0$ has a BR about
$10^{-3}$ ($2\times 10^{-5}$), for $m_{H^+}=300$ GeV and $\tan\beta=7$ (30).
Similarly, the BR for the mode $H^+ \to W^+Z$ is about $4 \times 10^{-2}$ ($4 \times 10^{-3}$), whereas the BR for $H^+ \to W^+\gamma$ is about $2 \times 10^{-6}$ ($2\times 10^{-7}$), for the same values of parameters. In this case, $H^+\to W^+Z$ dominates. 

{\bf c) Non-decoupling scenario-B.} Here we also assume a large mass difference between $A^0$ and $H^+$, i.e. $m_A-m_{H^+}=300$ GeV, with $m_H \simeq m_{H^+}$,
but now with $\alpha \simeq \beta-\pi/4$.
In this case we find that the BR for 
the mode $W^+h^0$ has a BR about
$1$ ($0.2$) for $m_{H^+}=300$ GeV and $\tan\beta=7$ (30).
Similarly, the BR for the mode $H^+ \to W^+Z$ is about
$ 10^{-2}$ ($4 \times 10^{-3}$), whereas
the BR for $H^+ \to W^+\gamma$ is about
$4\times 10^{-7}$ ($ 2\times 10^{-7}$), for the same values of parameters. In this scenario $H^+\to W^+h^0$ clearly dominates, even above the $t\bar{b}$ mode.

{\bf{ 3.- Charged Higgs decays in the effective Lagragian approach.}} 
The relevant  operators needed for an effective lagrangian
description of the THDM were previously studied in ref.\cite{wudkamexico}.
Here we show those operators of the lower dimension that could potentially
induce corrections to the charged Higgs vertices. The corresponding
effective Lagrangian has the following form:
\begin{eqnarray}
\mathcal{L}_{eff} = \mathcal{L}_{THDM} + \sum_{n \geq 6}\bigg[\sum_i 
\frac{\alpha^i_n}{\Lambda^{n-4}}(\mathcal{O}^i_n+h.c.)\bigg]
\end{eqnarray} 
where $\mathcal{L}_{THDM}$ is  the  THDM Lagrangian, $\Lambda$
is the new physics scale, $\mathcal{O}^i_n$ are higher--dimension
$SU(2)\times U_Y(1)$ invariant operators, and $\alpha^i_n$ are
unknown parameters, whose order of magnitude can be estimated
because gauge invariance makes possible to establish the order
of perturbation theory at which each operator can be generated
in the fundamental theory \cite{AEW}. This fact allows to
introduce a hierarchy among the operators of a given
dimensionality, which has important consequences from the practical
point of view, since the operators generated at loop levels
would be suppressed by the loop factor $(4\pi)^{-2}$ with respect
to those induced at tree level. In the following we shall consider
only the lower dimension operators, namely, those of dimension--six 
that may be generated at tree level in the fundamental theory.
It will be seen below, after SSB some of these operators introduce
modifications to the quadratic terms of the dimension--four theory, 
so a redefinition of fields and parameters will be needed
in order to obtain the canonical form for the propagators.
 We shall classify the operators that modify the charged Higgs vertices,
and could potentially induce corrections to the 
charged Higgs decays $H^+ \to W^+V$ and $H^+ \to W^+h^o$,
as follows:

{\bf {\it i}) Operators that contain only scalar fields.}
These operators modify the Higgs potential and the kinetic structure of
the scalar fields, and are given by:
\begin{eqnarray}
\mathcal{O}^{\phi}_{ijklmn}& = & \frac{1}{3} (\Phi_{i}^{\dag } \Phi_{j} )
(\Phi_{k}^{\dag } \Phi_{l})(\Phi_{m}^{\dag } \Phi_{n}) \label{i1} \\
\mathcal{O}^{\partial \phi}_{ijkl} & = & \frac{1}{2} 
\partial_{\mu}( \Phi_{i}^{\dag } \Phi_{j})
\partial^{\mu} ( \Phi_{k}^{\dag } \Phi_{l}) \label{i2}
\end{eqnarray}
where $i,j,k,l,m,n=1,2$ and $\Phi_i$ is a Higgs doublet.
There are several operators, depending on the possible
combinations $i,j,k,l$ of the two doublets, but for our purpose,
it is sufficient to consider only those that satisfy the discrete
symmetry $\Phi_1 \to \Phi_1$ and $\Phi_2 \to -\Phi_2$.
This symmetry allows only eight operators of the type (\ref{i1}),
from which the combinations $(111212),(221212),(111221)$, and
$(221221)$ do not respect the custodial symmetry.
On the other hand, there are five operators of the type
(\ref{i2}) and two of them violate the custodial symmetry
too, namely those corresponding to the combinations 
$(1221)$ and $(1212)$. \\

When the operators (9) are included in the Higgs potential,
they will induce
modifications to the  minimization conditions,
and the angle $\alpha $ (but not $\beta $) need to be
redefined. This redefinition can be worked out to the
first order in the parameters $\alpha^i$. Moreover, the operators 
(9,10), when  included  in the Higgs sector of model, 
will modify the Feynman rules for the scalar couplings,
and will induce new contributions to
the loop-amplitude for the  decay $H^+\to W^+\gamma$ 
through the vertices $H^+H^-\phi$. 
These operators also affect the mass relations
of the physical Higgs states as follows
\begin{eqnarray}
m_\phi^2=m_{\phi}^{\circ 2}(1+\sum_i \epsilon^i_{\phi})
\end{eqnarray}
where  $\phi=H^+ ,H^o ,h^o, A^o$ and the parameters
$\epsilon_{\phi}^i $ are functions of the ratio of
energy scales $(v/\Lambda)^2$ and the parameters $\alpha^i $, 
associated with the contributions from the operators 
$\mathcal{O}^i$ (in what follows, it will 
be understood that the coefficient $\epsilon$ contain the multiplicative factor $(v/\Lambda)^2$).
$m^\circ_{\phi}$ denotes the mass of the $\phi$ state within THDM, i.e., the superscript $\circ$ denotes mass relations arising from the dimension--four theory.\\

{\bf {\it ii)} Operators that modify the Higgs-gauge boson interactions.}
This set includes the following,
\begin{eqnarray}
\mathcal{O}_{ijkl}^{(1)} & = & \Phi^{\dag }_{i} \Phi_{j} (D_{\mu} \Phi_{k})^{\dag }
(D^{\mu} \Phi_{l}) \label{ii1}\\
\mathcal{O}_{ijkl}^{(2)} & = & \left[ \Phi^{\dag }_{i} (D_{\mu} \Phi_{j}) \right] 
\left[ (D_{\mu} \Phi_{k})^{\dag } \Phi_{l} \right] \label{ii2}
\end{eqnarray}
There are six operators of each type, consistent with the
discrete symmetry $\Phi_1\to \Phi_1, \Phi_2\to -\Phi_2$. 
While all six operators of the type
(\ref{ii2}) violate the custodial symmetry, the ones of the type
(\ref{ii1}) do respect it. In fact, these operators
lead to a redefinition of the gauge boson masses given by
\begin{eqnarray}
m^2_W&=&m^{\circ 2}_W(1+\epsilon^{(1)}_z),\\
m_Z^2&=&m^{\circ 2}_Z(1+\epsilon^{(1)}_z + \epsilon^{(2)}_z),
\end{eqnarray}
where $\epsilon^{(1)}_z$ and $\epsilon^{(2)}_z$ are the
modifications introduced by the operators (\ref{ii1}) and
(\ref{ii2}) to the gauge boson masses, respectively, 
which imply the constraint
\begin{equation}
\Delta \rho = | \epsilon_{z}^{(2)}| \leq 0.003
\end{equation}

This shows that operators (\ref{ii2})
do not respect the custodial symmetry, which automatically guarantee
the existence of the $H^+W^-Z$ vertex, but this will be
suppressed by the $\rho$ parameter. However, it is important to notice
that both operators (\ref{ii2}) and (\ref{ii1}) induce the 
$H^+W^-Z$ vertex, so the effective Lagrangian approach
tells us that this interaction can be directly induced by operators
that respect the custodial symmetry, whose contributions to this
vertex may be dominant since they are generated at tree level in
the fundamental theory. Thus, we can study the decay
$H^+\to W^+Z$ considering only the vertex contributions arising from the
set of operators (\ref{ii1},\ref{ii2}), including as well the loop
prediction of the THDM. On the other hand, the contribution of this vertex 
to the decay $H^+\to W^+\gamma$ should also be considered since it is a
loop contribution in the context of the full theory.\\

Moreover, the operators (\ref{i1},\ref{i2},\ref{ii1},\ref{ii2}) modify
the prediction of the dimension--four theory for the relation
between the masses of the charged and the neutral CP--odd Higgs, as follows 
\begin{equation}
m_{H^{\pm}}^{2}=m_{A^{o}}^{2}+\frac{2m_{W}^{2}}{g^{2}}\bigg[
(1-\epsilon^{(1)}) 
(\lambda_{5}-\lambda_{4})
-\epsilon_{AH}^{\phi}+\frac{g^{2}m_{A^{o}}^{2}}{2m_{W}^{2}}(\epsilon_{A^{o}}^{\partial \phi}+\epsilon_{A^{o}}^{(2)})\bigg],
\end{equation}
where $\epsilon^{(1)}$ denote a combination of the
coefficients of the operators (\ref{ii1}), $\epsilon_{AH}^{\phi}$
are contributions of the operators (\ref{i1}), which violate the
custodial symmetry. $\epsilon^{\partial \phi}_{A^0}$ and
$\epsilon^{(2)}_{A^0}$ are functions of the coefficients
of the operators (\ref{i2}) and (\ref{ii2}), respectively,
which do not respect the custodial symmetry neither.
When $\lambda_4=\lambda_5$ and
$\epsilon_{A^o}^{\partial \phi}, \epsilon^{(2)}_{A^0},
\epsilon_{AH}^\phi \to 0 $
\footnote{By inspecting the corresponding expressions for
the $\epsilon$'s, one could see that those
cancellations associated with the $SU_C(2)$ limit, indeed appear.},
we have $m_{H^\pm}=m_{A^o}$, and the custodial
symmetry of the Higgs potential is recovered.\\

{\bf {\it iii}) Yukawa-like operators.} This set will involve the
fermion fields, but here we shall not consider fermion mixing and
will keep only the expressions for the 3rd family quarks, which dominates
the fermionic contribution to the loop amplitudes in $H^+ \to W^+\gamma $.
We include, 
\begin{eqnarray}
\mathcal{O}_{ijk}^{t \phi} & = & 
\Phi_{i}^{\dagger } \Phi_{j} \bar{Q}_{L} \tilde{\Phi}_{k} t_{R}
\\ 
\mathcal{O}_{ijk}^{b \phi} & = & \Phi_{i}^{\dagger } \Phi_{j} \bar{Q}_{L}\Phi_{k} b_{R},
\end{eqnarray}
where $Q_L=(t,b)_L$ is the left--handed $SU_L(2)$ doublet. After SSB, these operators introduce modifications into the fermion masses and also effective interactions of dimension four among the charged Higgs and the members of the 3rd family quarks. Thus, they contribute to $H^+\to W^+\gamma$ through the $H^+tb$ vertex. Since this vertex has a renormalizable structure, its contribution will be finite.

{\bf {\it iv}) Operators that will induce corrections to the charged
and neutral currents of the SM.} Here we consider the following 
operators,
\begin{eqnarray}
O_{ij}^{\phi Q (1)} & = & i (\Phi^{\dag}_{i}D_{\mu} \Phi_{j}) \bar{Q}_{L} 
\gamma^{\mu}Q_{L} \\ 
O_{ij}^{\phi Q (2)} & = & i (\Phi^{\dag}_{i}D_{\mu} \tau^{a} \Phi_{j}) 
\bar{Q}_{L}  \gamma^{\mu} \tau^{a} Q_{L} \\ 
O_{ij}^{\phi t} & = & i (\Phi^{\dag}_{i}D_{\mu} \Phi_{j}) \bar{t}_{R} 
\gamma^{\mu}t_{R} \\ 
O_{ij}^{\phi b} & = & i (\Phi^{\dag}_{i}D_{\mu} \Phi_{j}) \bar{b}_{R} 
\gamma^{\mu}b_{R}
\end{eqnarray}
These operators introduce modifications into the
dimension--four vertices $H^+tb$ and $W^+tb$,
which contribute to the loop decay $H^+\to W^+\gamma$.
Nonrenormalizable vertices of dimension--five of the type
$H^+tb\gamma$, $H^+W^-tt$, and $H^+W^-bb$ (and also the $H^+tb$ one with derivative structure) are also induced from operators (20-23). These dimension--five vertices lead to a divergent contribution to the loop-amplitude for the decay $H^+\to W^+\gamma$, which we have renormalized using the $\overline{MS}$ scheme \cite{AEWG}. This set of operators has the advantage that can be bounded by their contributions to
the S,T,U parameters, through the decay $\Gamma(Z\to \bar{b}b)$ \cite{KR}. Working at tree--level, as it was done in ref.\cite{SCW} for the analogous contribution of the effective operators of the minimal SM to the S, U, T parameters, we derive an order of magnitude bound, $\epsilon_i \leq O(10^{-3})$, which we will assume hereafter for the coefficients of all operators. \\

Now we can discuss the effect of these operators on the
charged Higgs interactions. In order to probe these effects 
one can first study the corrections to the 
decay width for $H^+ \to W^+h^o $; the result is given now 
as follows:
\begin{equation}
\Gamma(H^+\to W^+h^0)= \Gamma_{\tiny THDM} (H^+\to W^+h^0)
\left[ 1 + 
\frac{ \epsilon^{(1)}_{HWh}+ \epsilon^{(2)}_{HWh}}{c_{\beta - \alpha} }
\right]^2
\end{equation}
where $\Gamma_{\tiny THDM}(H^+\to W^+h^0)$ is given  in eq.(3) and
$\epsilon^{(1),(2)}_{HWh}$ includes effects from operators
(\ref{ii1},\ref{ii2}). Since the operators (\ref{ii2}) are highly
constrained by the custodial symmetry, this decay would be dominated
by the contributions arising from operators (\ref{ii1}).\\

Regarding the decay  $H^+\to W^+  Z$, we shall only consider the 
operators (\ref{ii1},\ref{ii2}), which induce effectively 
this vertex.
The resulting effective amplitude  will be added to the loop
result obtained in the previous section. The new form can be
obtained simply by replacing: $F_Z \to F_Z+\Delta F_Z$,
$G_Z \to  G_Z+\Delta G_Z$ and $H_Z \to H_Z+\Delta H_Z$,
with: $\Delta F_Z= \frac{S_{eff}}{c_w}$, $\Delta G_Z= \Delta H_Z= 0 $, and $S_{eff} = s_w^2 \epsilon_{HWZ}^{(1)} + \epsilon^{(2)}_{HWZ}$. The parameters $\epsilon^{(1,2)}_{HWZ}$  are functions of
coefficients that come from the operators (\ref{ii1},\ref{ii2}).
The operators (\ref{ii2}) are associated with the breaking of the
custodial symmetry in the Higgs sector, and the decay
$H^+ \to W^+ Z$ will be sensitive to its strength.  
 
Finally, for the decay $H^+ \to W^+\gamma$ we shall consider the
one-loop contribution induced by the tree--level--generated
dimension--six operators listed in the paragraphs
{\bf {\it i}), {\it ii}), {\it iii}), {\it iv})}; the operators of {\bf {\it iv})}
have the convenience of being constrained by the S,T,U parameters.
We have performed the loop calculation using a nonlinear gauge--fixing
procedure. This scheme is suited to eliminate vertices of the type $WVG$
involving the Goldstone bosons, not only from the renormalizable
Lagrangian but also from the effective operators (\ref{ii2}) that
introduce modifications to the unphysical vertex $WZG_W$. We have
introduced the following nonlinear gauge--fixing functions which
transform covariantly under the $U_e(1)$ group:
\begin{eqnarray}
f^{+} & = & \left[ D^e_{\mu }+\frac{igs^2_w}{c_w}\bigg(1+\frac{\epsilon^{(2)}_z}{s^2_w}\bigg)Z_\mu \right] W^{+\mu} 
- i \xi m_{W} G_{W}^{+}, \nonumber \\
f^{Z} & = & \partial _{\mu } Z^{ \mu }-  \xi m_{Z} G_{Z}^{o}, \\ 
f^{A} & = & \partial _{\mu } A^{ \mu }, \nonumber 
\end{eqnarray}
where $D^e_\mu$ is the electromagnetic covariant derivative
and $\xi$ is the gauge parameter. The Feynman Rules for vertices of
the type $WWZ$ are  modified too, and are consistently used to
obtain the loop amplitude (the full list of Feynman rules in this gauge as well
as  the corresponding graphs for the radiative decays will appear
elsewhere \cite{ournext} ).\\

At one--loop level in the full theory, the decay $H^+\to W^+\gamma$ also receives a direct contributions from loop--generated dimension--six operators. These operators are
\begin{eqnarray}
\mathcal{O}_{ij}&=&(\Phi^\dag_i{\bf W}_{\mu \nu}\Phi_j)B^{\mu \nu}, \\
\widetilde{\mathcal{O}}_{ij}&=&(\Phi^\dag_i{\bf W}_{\mu \nu}\Phi_j)\widetilde{B}^{\mu \nu},
\end{eqnarray}
where ${\bf W}_{\mu \nu}=\tau^a W_{\mu \nu}$ and $\widetilde{B}_{\mu \nu}=\frac{1}{2}\epsilon_{\mu \nu \lambda \rho}B^{\lambda \rho}$. In the calculation of this decay we have explicitly introduced the loop factor $(4\pi)^{-2}$ in the coefficients of these operators. \\

Finally, the expressions for the decay width of $H^+\to W^+\gamma$ can be written  as in  eq.(5), by making the substitutions, $G_\gamma \to G_\gamma+\Delta G_\gamma$ and $H_\gamma \to H_\gamma+\Delta H_\gamma$. Besides the amplitudes induced by the above loop--generated operators, the terms $\Delta G_\gamma$ and $\Delta H_\gamma$ contain the loop contributions coming from the tree--level--generated operators given in paragraphs {\bf {\it i}), {\it ii}), {\it iii}), {\it iv})}. The result depends on Passarino--Veltman scalar two-- and three--point functions, and due to the presence of nonrenormalizables vertices induced by the operators of {\it iv)}, the divergent parts of some two--point functions do not disappear, so a renormalization scheme must be adopted. We used the $\overline{MS}$ scheme with the renormalization scale specified by $\mu=\Lambda$, which leads to a logarithmic dependence in the corresponding amplitudes of the form $ln(m^2_i/\Lambda^2)$, $m_i$ being one of the masses circulating in the loops. In order to evaluate these logarithms, we have estimated the value of $\Lambda$ using the bounds obtained from the S, T, U parameters. The expressions for these contributions are very long too and will be presented elsewhere  \cite{ournext}.\\

In the description of physics beyond the Fermi scale that we are
using, the corrections coming from the higher-dimensional operators
can represent new physics of perturbative type, e.g. SUSY particles,
new fermions, etc, whose coefficients will be typically small
($ \alpha_{i} \approx 0.1-1$). Assuming that new physics would
be apparent at scales of order $\Lambda \approx 1$ TeV, a typical
tree--level--generated dimension--six operator will have a
suppression factor within the range
$(v/\Lambda)^2\alpha_i \approx 10^{-2}-10^{-1}$. However,
in order to make predictions, we have adopted a more conservative
point of view by choosing values for the coefficients of the various
operators of order $10^{-3}$, similar to the bounds determined for the operators that are constrained by the S, T, U parameters. The numerical results are shown in Figs.1,2,3. We comment here on the changes induced by the effective operators for the charged Higgs branching ratios, of the three scenarios discussed in the THDM section, taking the same values of parameters that  define each case, namely

{\bf a) SUSY-like scenario.} 
For $tan\beta=7$, the BR for the mode $W^+h^0$ goes from
$ 10^{-2}$ (THDM)  up to $ 2\times 10^{-2}$ (Eff. Lagrangian);
whereas the BR for the mode $H^+ \to W^+Z$ goes from
$10^{-3}$ (THDM) to $7 \times 10^{-3}$(Eff. Lagrangian), and
 $H^+ \to W^+\gamma$ goes from $2\times 10^{-6}$ (THDM) up to
$ 10^{-5}$ (Effective Lagrangian), which is the mode with largest
enhancement for this scenario.

\begin{figure}
\centering
\includegraphics[width=5in]{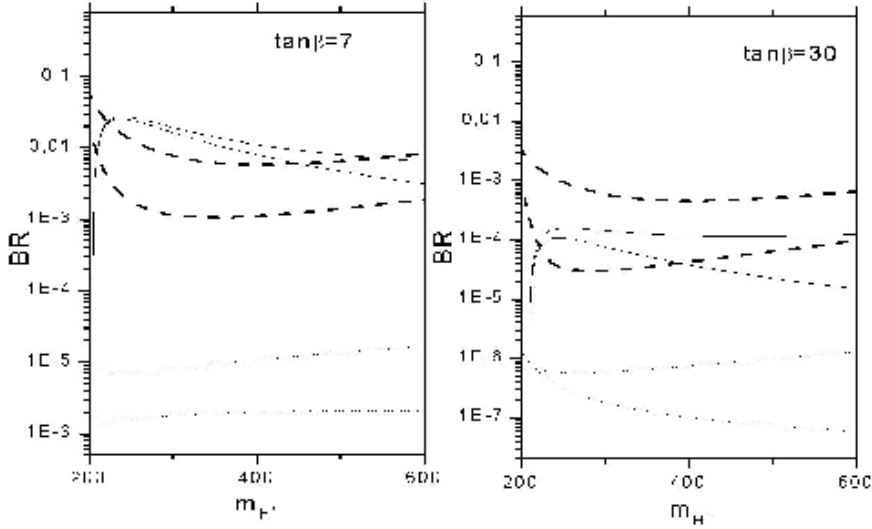}
\caption{ Branching ratios for charged Higgs decay into $Wh$ (solid),
$WZ$ (dashes) and $ W \gamma$ (dots) for the parameters, $m_{h}=115$ GeV,
$m_{H^{\pm}} \simeq m_{A} \simeq m_H $ and $\alpha \simeq \beta - \pi/2$.
Lower (upper) lines correspond to the THDM (effective Lagrangian) cases.}
 \end{figure}

{\bf b) Non-decoupling scenario-A.} In this case, for $\tan\beta=7$ again, the BR for the mode $W^+h^0$ goes from
$ 10^{-3}$ (THDM)  up to $ 2\times 10^{-3}$ (Eff. Lagrangian);
whereas the BR for the mode $H^+ \to W^+Z$ goes from
$4 \times 10^{-2}$ (THDM) to $ 10^{-1}$(Effective Lagrangian), and
 $H^+ \to W^+\gamma$ goes from $2\times 10^{-6}$ (THDM) up to
$10^{-5}$ (Effective Lagrangian).

\begin{figure}
\centering
\includegraphics[width=5in]{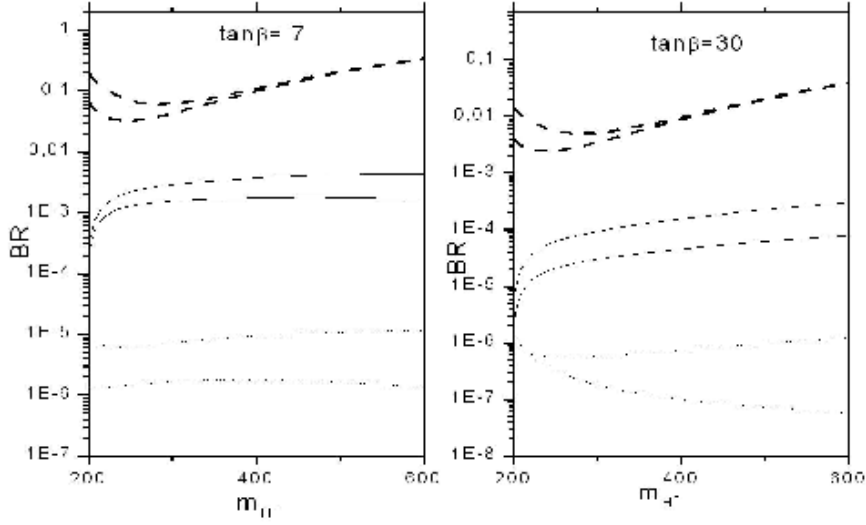}
\caption{ Same as in fig. 1, but now with  parameters  
$m_{H^{\pm}} - m_{A}=300 $ Gev, $m_H \simeq m_{H^{\pm}}$ 
and $\alpha \simeq \beta - \pi/2$.}
\end{figure}

{\bf c) Non-decoupling scenario-B.} For $\tan\beta=7$, the BR for the mode $W^+h^0$ remmains approximately constant for both the THDM and the Eff. Lagrangian; whereas the BR for the mode $H^+ \to W^+Z$ goes from $10^{-2}$ (THDM) to $2 \times 10^{-2}$(Effective Lagrangian), and
 $H^+ \to W^+\gamma$ goes from $3\times 10^{-7}$ (THDM) up to $ 10^{-6}$ (Eff. Lagrangian).
 
Results for $\tan\beta=30$ show a similar behaviour for all previous cases. 

\begin{figure}
\centering
\includegraphics[width=5in]{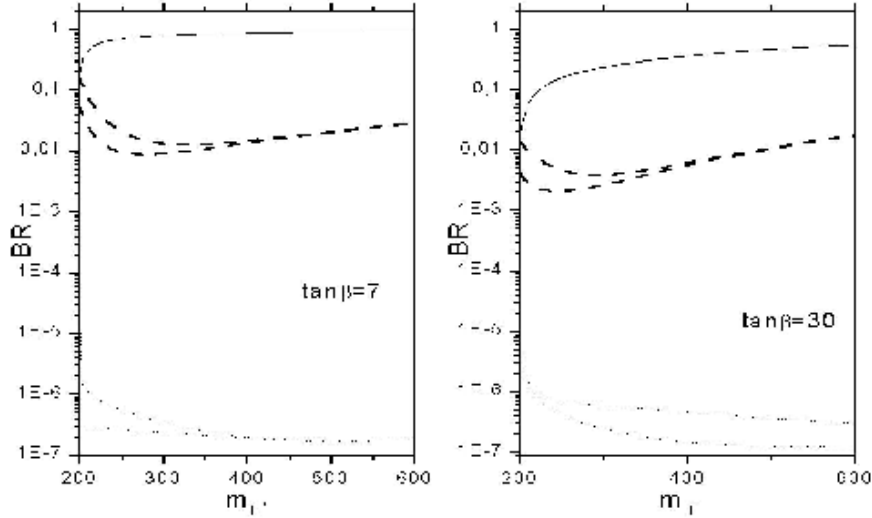}
\caption{Same as in fig. 1, but now with  parameters
$m_{H^{\pm}} - m_{A}=300 $ Gev, $m_H \simeq m_{H^{\pm}}$ 
and $\alpha \simeq \beta - \pi/4$.  }
\end{figure}

{\bf{4.- Conclusions.}}
We have studied the rare decays of the charged
Higgs boson $H^+ \to W^+ h^o, W^+ Z, W^+ \gamma $
as possible tests of the symmetries of the Higgs sector.
Starting from the two-higgs doublet model, where the radiative
decays are in general suppressed, we construct an effective 
Lagrangian extension of the model, that describes the modified 
charged Higgs interactions.
The S,T,U parameters are used to impose bounds on certain
dimension--six operators that describe some of these interactions, and 
used to make predictions on the decays $H^+ \to W^+ \gamma, W^+Z, W^+h^0$. 
We find that theses modes can receive an enhancement that could be tested at future colliders.
For the discussion of results we have identified three
scenarios, whose characteristics can be tested at future
colliders, as they lead to different predictions for these
modes, namely: a) a SUSY-like case, where the modes $WZ$ and $Wh$ have
similar BR, b) a non-decoupling scenario, with parameters leading
the $WZ$ mode to become the dominant one,
 and c) a non-decoupling case where the mode $Wh$
becomes the dominant one. As we can see from Figures 1--3, there is a large enhancement for the modes $W^+\gamma$ and $W^+Z$, when passing from the THDM to the effective Lagrangian framework, whereas the mode  $W^+h^0$ receives a moderate enhancement.

\bigskip

{\bf{Acknowledgments}.} We acknowledge financial support from CONACYT and
SNI (M\' exico). Fruitful discussions with J. M. Hern\' andez and M.A. P\' erez are also acknowledged.

\end{document}